\documentclass[aps,prl,floatfix,twocolumn,superscriptaddress]{revtex4}

\usepackage{graphics}
\usepackage{epsfig}

\begin{document}

\title{Accurate determination of tensor network state of quantum
lattice models in two dimensions}

\author{H. C. Jiang}
\author{Z. Y. Weng}
\affiliation{Center for Advanced Study, Tsinghua
University, Beijing, 100084, China}

\author{T. Xiang}
\affiliation{Institute of Physics, Chinese Academy of Sciences, P.O.
Box 603, Beijing 100190, China}

\affiliation{Institute of Theoretical Physics, Chinese Academy of
Sciences, P.O. Box 2735, Beijing 100190, China}

\date{\today}

\begin{abstract}
We have proposed a novel numerical method to calculate accurately
the physical quantities of the ground state with the
tensor-network wave function in two dimensions. We determine the
tensor network wavefunction by a projection approach which applies
iteratively the Trotter-Suzuki decomposition of the projection
operator and the singular value decomposition of matrix. The norm
of the wavefunction and the expectation value of a physical
observable are evaluated by a coarse grain renormalization group
approach. Our method allows a tensor-network wavefunction with a
high bond degree of freedom (such as $D=8$) to be handled
accurately and efficiently in the thermodynamic limit. For the
Heisenberg model on a honeycomb lattice, our results for the
ground state energy and the staggered magnetization agree well
with those obtained by the quantum Monte Carlo and other
approaches.
\end{abstract}


\maketitle

The application of the density matrix renormalization group (DMRG)
proposed by White\cite{White1992} has achieved great success in
one dimension in both zero and finite
temperatures\cite{Bursill96,Wang97,Shollwock}. However, in two
dimensions, the application of the DMRG in both the real and
momentum space\cite{Xiang01,Xiang96} has been limited only to
small lattices. The error resulting from the DMRG truncation
increases extremely fast with increasing size of lattice. To
resolve this problem, the tensor-network state, which is an
extension of the matrix product in one
dimension\cite{Ostlund1995}, was proposed\cite{Verstraete}. In a
tensor network state, a local tensor is directly entangled with
the other local tensors on all the directions of the lattice. This
leads to two problems in the treatment of the tensor network
state. First, it is difficult to determine accurately all the
elements of local tensors by any variational approach since the
total degree of freedom of a local tensor increases exponentially
with the dimension of the tensor. Second, it is even more
difficult to calculate the expectation value of any physical
observable even if we know the expression of the tensor network
wave function, since the number of summations over the basis
configurations increases exponentially with lattice size.

In this paper, we propose a novel method to handle the
tensor-network wave function in two dimensions. We will show that
the tensor network wavefunction $|\Psi\rangle$ of the ground state
can be accurately determined by applying an iterative projection
approach. This approach is similar to the time-evolving block
decimation method that was used to determine the matrix product
wave function of the ground state in one dimension\cite{Vidal2003,
Vidal2007} Then we will generalize the classic coarse grain
renormalization group approach proposed by Levin and
Nave\cite{Levin2007} to the quantum system, and use it to
calculate the norm of the wave function and the expectation value
of any physical observable. This provides an accurate and
efficient tool to determine the expectation values of physical
quantitise from the tensor-network wavefunction of the ground
state.

\begin{figure}[tbp]
\centerline{
    \includegraphics[height=2.4in,width=3.4in]{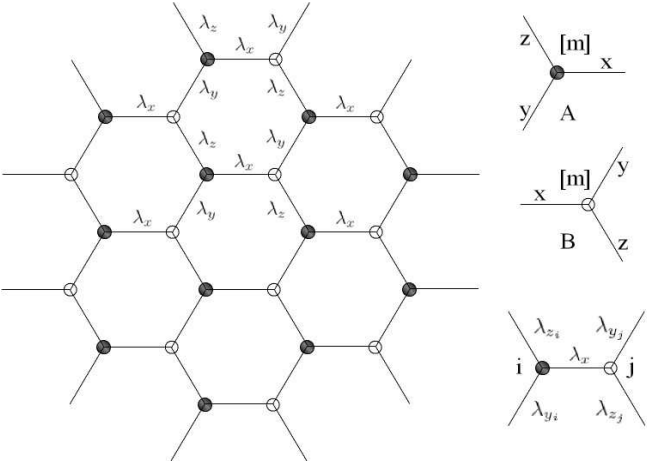}
    }
\caption{ Schematic representation of a tensor network state on a
honeycomb lattice. The lattice is divided into two sublattices,
represented by the black and white dots, respectively. Each
vertex, on which a spin state is inhibited, is connected with
three neighboring vertices along three directions, labelled by
$x$, $y$, and $z$. On each bond, there is a diagonal matrix (or a
vector), $\lambda_\alpha$, where the subscript $\alpha = x$, $y$
or $z$ is a bond index of dimension D. At each vertex, a tensor
representation of the spin state $m$, $A_{x,y,z} [m]$ for the
black sublattice or $B_{x,y,z} [m]$ for the white sublattice, is
defined. A tensor network state \label{eq:tensor} is a product of
all these bond vectors and vertex tensors. } \label{fig:tns}
\end{figure}

Below, we will take the S=1/2 Heisenberg model on a honeycomb
lattice
\begin{eqnarray}
H & = & \sum_{\langle ij \rangle} H_{ij}, \\
H_{ij} & = & J  S_i \cdot S_j - \frac12 h \left[ (-)^i S_{i,z}+
(-1)^j S_{j,z} \right] , \label{eq:model}
\end{eqnarray}
as an example to show how the method works. Here, $\langle i j
\rangle$ means that $i$ and $j$ are the two nearest neighboring
sites, $h$ is the magnitude of a staggered magnetic field. It is
straightforward to extend the method to other quantum lattice
models with short range interactions in two dimensions. We assume
the tensor network state to have the following form
\begin{eqnarray}
|\Psi\rangle &=& \mathrm{Tr} \prod_{i\in b, j\in w } \lambda_{x_i}
\lambda_{y_i} \lambda_{z_i} A_{x_i y_i z_i}
[m_i]  B_{x_j y_j z_i} [m_j]\nonumber \\
&& |m_i m_j \rangle. \label{eq:tns}
\end{eqnarray}
A schematic representation of this tensor network state is shown
in Fig. \ref{fig:tns}.  In Eq. (\ref{eq:tns}), '$b/w$' stands for
the black/white sublattice. $m_i$ is the eigenvalue of $S_{iz}$.
$A_{x_i y_i z_i} [m_i]$ and $B_{x_j y_j z_j} [m_j]$ are the two
three-indexed tensors defined on the black and white sublattices,
respective. $\lambda_{\alpha_i}$ ($\alpha = x, y, z$) is a
positive diagonal matrices (or vectors) of dimension D defined on
the bond emitted from site $i$ along the $\alpha$ direction. The
subscripts $x_i$, $y_i$ and $z_i$ are the integer bond indices of
dimension D (i.e. each running from $1$ to $D$). A bond links two
sites. The two bond indices defined from the two end points take
the same values. For example, if the bond connecting $i$ and $j$
along the $x$ direction, then $x_i=x_j$. The trace is to sum over
all spin configurations $\{ \cdots, m_i, m_j, \cdots\}$ and over
all bond indices.

The ground state wavefunction can be determined by applying the
projection operator $\exp (-\tau H)$ to an arbitrary initial state
$|\Psi \rangle$. In the limit $\beta \rightarrow \infty$, $\exp
(-\tau H) |\Psi \rangle$ will converge to the ground state of $H$.
However, this projection cannot be done in a single step since the
terms in $H$ defined by Eq. (\ref{eq:model}) do not commute with
each other. In real calculation, we will take a small $\tau$ and
apply this projection operator to $|\Psi\rangle$ iteratively for
many times.

Let us start by dividing the Hamiltonian into three parts
\begin{eqnarray}
H & = & H_x + H_y + H_z ,\nonumber \\
H_\alpha & = & \sum_{i\in \mathrm{black} } H_{i,i+\alpha} \quad
(\alpha = x, y, z). \nonumber
\end{eqnarray}
$H_\alpha$ ($\alpha = x, y, z$) contains all the interaction terms
along the $\alpha$-direction only. These terms commute with each
other. From the Suzuki-Trotter formula, we can then express the
projection operator as
\begin{eqnarray}
e^{-\tau H} \approx  e^{-\tau H_z} e^{-\tau H_y} e^{-\tau H_x} +
o(\tau^2).
\end{eqnarray}
This means that each iteration of projection can be done using
$\exp (-\tau H_\alpha)$ ($\alpha = x, y, z$) in three separate
steps.

In the first step, the projection is done with $H_x$. As only the
two neighboring spins connected by horizontal bonds have
interactions in $H_x$, the resulting projected wavefunction can
expressed as
\begin{eqnarray}
&& e^{-\tau H_x}|\Psi \rangle \nonumber \\
&=& \mathrm{Tr} \prod_{i\in b, j=i+x} \sum_{m_i m_j}   \langle
m_i^\prime m_j^\prime | e^{-H_{ij}\tau} |m_i m_j\rangle
\nonumber \\
&& \lambda_{x_i}\lambda_{y_i} \lambda_{z_i} A_{x_i y_i z_i} [m_i]
B_{x_j y_j z_j} [m_j] |m_i^\prime m_j^\prime \rangle.
\end{eqnarray}
From this, a $(D^2d)\times (D^2d) $ matrix can be defined by
\begin{eqnarray}
&&S_{y_i z_i m_i^\prime  ,  y_j z_j m_j^\prime}\nonumber \\
& = & \sum_{m_i m_j}\sum_x \langle m_i^\prime m_j^\prime |
e^{-H_{ij}\tau}| m_i
m_j\rangle\nonumber \\
&& \lambda_{y_i} \lambda_{z_i} A_{x y_i z_i}[m_i] \lambda_x B_{x
y_j z_j} [m_j] \lambda_{y_j} \lambda_{z_j} , \label{eq:S1}
\end{eqnarray}
where $d=2$ is the total number of states of a S=1/2 spin. Taking
the singular value decomposition for this matrix, one can further
express this $S$ matrix as
\begin{equation}
S_{y_i z_i m_i  ,  y_j z_j m_j} = \sum_x U_{y_i z_i m_i, x}
\tilde{\lambda}_x V^T_{x, y_j z_j m_j} \label{eq:svd}
\end{equation}
where $U$ and $V$ are two unitary matrices and $\tilde{\lambda}_x$
is a positive diagonal matrix of dimension $D^2d$.

Next we truncate the basis space by keeping only the $D$ largest
singular values of $\tilde{\lambda}_x$. Then we set the left
$\tilde{\lambda}_x$ as the new $\lambda_x$ ($x=1\cdots D$) and
update the tensors $A$ and $B$ by the following formula
\begin{eqnarray}
A_{x y_i z_i} [m_i] & = &  \lambda_{y_i}^{-1}
\lambda_{z_i}^{-1} U_{y_i z_i m_i, x}, \label{eq:A}\\
B_{x y_j z_j} [m_j] & = & \lambda_{y_j}^{-1} \lambda_{z_j}^{-1}
V_{y_j z_j m_j^\prime, x}. \label{eq:B}
\end{eqnarray}

\begin{figure}[tbp]
\centerline{
    \includegraphics[height=2.0in,width=2.8in]{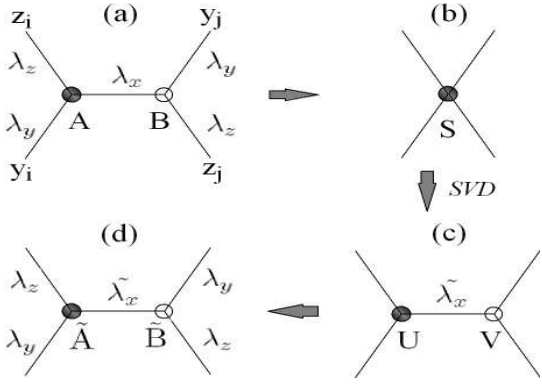}
    }
\caption{Flow chart of the one-step renormalization of the wave
function. (a) To use $\exp (-H_{i,i+x}\tau)$ to act on the tensor
network state. (b) To evaluate the $S$-matrix defined by Eq.
(\ref{eq:S1}). (c) Perform the singular value decomposition for
$S$. (d)Truncate the basis space of $\tilde{\lambda}_x$ and find
$\tilde{A}$ and $\tilde{B}$ with Eqs. (\ref{eq:A}) and
(\ref{eq:B}), respectively. \label{fig:evolution}}
\end{figure}

A flow chart of the above one-step renormalization of the wave
function is shown in Fig. \ref{fig:evolution}. The next two steps
of projections can be similarly done with $H_y$ and $H_z$,
respectively. This completes one iteration of the projection. By
repeating this iteration proceduce many times, an accurate ground
state wave function can then be projected out. This iteration
process is very efficient. The converging speed depends on the
truncation error. In our calculation, we take $\tau=10^{-3}$
initially and then gradually reduce it to $\sim 10^{-5}$ to ensure
the convergence of the wavefunction. The number of iterations used
in our calculation is generally around $10^5 \sim 10^6$.

Given $|\Psi\rangle$, the expectation value of a measurement
quantity $O$ is defined by
\begin{eqnarray}
\langle \hat{O}\rangle = \frac{\langle \Psi| \hat{O}|\Psi\rangle
}{\langle \Psi|\Psi\rangle}. \label{physquan}
\end{eqnarray}
We notice that both $\langle \Psi|\Psi\rangle$ and $\langle \Psi|
\hat{O}|\Psi\rangle$ are tensor-network functions. For example,
\begin{eqnarray}
\langle \Psi| \Psi\rangle = \mathrm{Tr} \prod_{i\in b, j\in w}
T_{x_ix^\prime_i,y_iy^\prime_i,z_iz^\prime_i}^a
T_{x_jx^\prime_j,y_jy_j^\prime, z_j z_j^\prime}^b ,
\end{eqnarray}
where the trace is to sum over all bond indices. Both $T^a$ and
$T^b$ are $D^2\times D^2 \times D^2$ tensors. $T^a$ is defined by
\begin{eqnarray}
T^a_{xx^\prime, yy^\prime, zz^\prime}  & =& \sum_{m} \left(
\lambda_x \lambda_y\lambda_z\right)^{1/2} A_{xyz}[m] \nonumber \\
&& A_{x^\prime y^\prime z^\prime}[m] \left( \lambda_x^\prime
\lambda_y^\prime \lambda_z^\prime \right)^{1/2}.
\end{eqnarray}
$T^b$ is similarly defined. Thus we can apply the tensor
renormalization group method proposed by Levin et
al.\cite{Levin2007} to evaluate $\langle \Psi|\Psi\rangle$ and
$\langle \Psi| \hat{O}|\Psi\rangle$.

\begin{figure}[tbp]
\centerline{
    \includegraphics[height=1.0in,width=3.2in]{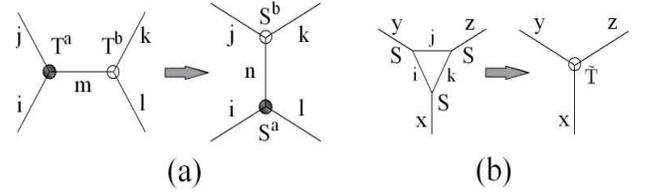}
    }
\caption{Steps of coarse-grain: (a) Form the $M$ matrix by tracing
out the common bond indices of tensors $T^a$ and $T^b$ defined on
the two neighboring sites with Eq. (\ref{eq:M}), and then perform
the singular value decomposition and find two new tensors $S^a$
and $S^b$ defined by Eqs. (\ref{eq:Sa}) and (\ref{eq:Sb}). (b)
Trace out all common bond indices of $S^a$-tensors (similarly for
the $S^b$-tensors) on a triangle formed by the three closed
vertices to form a coarse-grained tensor $\tilde{T}^a$ defined by
Eq. (\ref{eq:tta}). } \label{fig:cg}
\end{figure}

To perform the tensor renormalization, we first take two $T^a$ and
$T^b$ on the two ends of a bond and define the following
$D^4\times D^4$ matrix
\begin{equation}
M_{ll^\prime, kk^\prime} = \sum_n T^a_{nl^\prime k} T^b_{nk^\prime
l}.
\end{equation}
By taking the singular value decomposition, one can also express
this matrix as
\begin{equation}
M_{ll^\prime,kk^\prime} = \sum_{n=1\cdots D^4} U_{ll^\prime,n}
\Lambda_n V_{kk^\prime,n}, \label{eq:M}
\end{equation}
where $U$ and $V$ are unitary matrices, $\Lambda_n$ is a positive
defined diagonal matrix of dimension $D^4$. Again we will truncate
the basis space and keep only the basis states corresponding to
the largest $D^2$ singular values of $\Lambda$. Then the $M$
matrix can be approximately expressed as
\begin{eqnarray}
M_{ll^\prime,kk^\prime} \approx \sum_{n=1\cdots D^2}
S^a_{nll^\prime} S^b_{nkk^\prime} ,
\end{eqnarray}
where
\begin{eqnarray}
S^a_{nll^\prime} &=& \sqrt{\Lambda_n} U_{ll^\prime ,n}, \label{eq:Sa} \\
S^b_{nkk^\prime} & = & \sqrt{\Lambda_n} V_{kk^\prime,n}
\label{eq:Sb}
\end{eqnarray}
are the two vertex tensors defined in the new lattice shown in
Fig. (\ref{fig:cg}a).

\begin{figure}[tbp]
\centerline{
    \includegraphics[height=1.6in,width=3.4in]{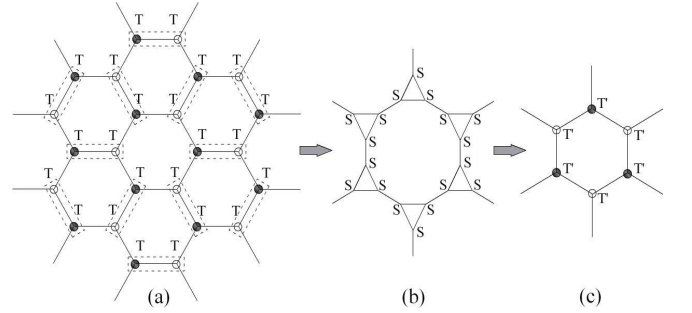}
    }
\caption{Tensor renormalization transformation on the honeycomb
lattice. } \label{fig:TRG}
\end{figure}

After the above transformation, the lattice structure is changed
(Fig. \ref{fig:cg}b). Now we replace each smallest triangle by a
single lattice point. This introduce a coarse-grained honeycomb
lattice with two coarse-grained tensors $\tilde{T}^{a}$ and
$\tilde{T}^{b}$ defined by
\begin{eqnarray}
\tilde{T}^a_{xyz} &=& \sum_{ijk} S^a_{xik} S^a_{yji} S^a_{zkj} \label{eq:tta} \\
\tilde{T}^b_{xyz} &=& \sum_{ijk} S^b_{xik} S^b_{yji} S^b_{zkj}.
\label{eq:ttb}
\end{eqnarray}
This coarse grain transformation reduces the lattice by a factor
of 3 at each iteration. Iterating this procedure, at the end the
honeycomb lattice will eventually becomes $6$ (Fig.
\ref{fig:TRG}). One can then trace out all bond indices to find
the norm of the wavefunction.

\begin{figure}[tbp]
\centerline{
    \includegraphics[height=2.2in,width=3.6in]{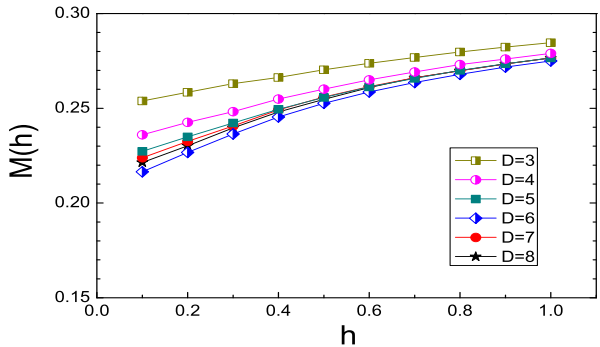}
    }
\caption{(color online) The staggered magnetization $M(h)$ as a
function of the staggered magnetic field, at different $D$. }
\label{fig:StaggerM}
\end{figure}

The above coarse grain tensor renormalization group transformation
can be straightforwardly extended to evaluate $ \langle \Psi|
\hat{O}|\Psi\rangle$. The difference is that $T^a$ and $T^b$ now
may become site dependent and their definitions are changed.

We have applied the above approach the spin-$\frac{1}{2}$
antiferromagnetic Heisenberg model (\ref{eq:model}). Both the
ground state energy and the staggered magnetization $M$ defined by
\begin{equation}
M(h) = \frac{E(h) - E_0}{h} \label{eq:stg}
\end{equation}
are calculated. In Eq. (\ref{eq:stg}), $E(h)$ is the ground state
energy in a finite staggered magnetic field $h$. The lattice size
is $N=6\times 3^{10}$. The finite size effect is negligible
compared with the truncation error resulted from the coarse grain
renormalization.

Table \ref{tab:energy} shows the ground state energy and the
staggered magnetization as a function of $D$ for the Heisenberg
model with $h=0$. The zero field staggered magnetization is
obtained by extrapolating $M(h)$ obtained at finite $h$ (Fig.
\ref{fig:StaggerM}) to the limit $h\rightarrow 0$. With $D=8$, we
find that the ground state energy $E = -0.5506$ and the staggered
magnetization $M=0.21 \pm 0.01$ in the zero field limit. They
agree well with the results obtained by the other approaches (see
Table \ref{tab:compare}).

In conclusion, we have proposed a novel method to treat the
tensor-network wave function of quantum lattice models in two
dimensions. It allows us to treat a tensor-network state with $D$
as high as 8 accurately and efficiently. The ground state energy
and the staggered magnetization of the S=1/2 Heisenberg model on
the honeycomb lattice obtained with this method are consistent
with those obtained by other methods.

We acknowledge the support of NSF-China and the National Program
for Basic Research of MOST, China.

\begin{table}[h]
  \centering
\caption{The ground state energy per site $E$ and the staggered
magnetization $M$ in the zero field limit as a function of $D$.}
  \vspace{2mm}
  \begin{tabular}[b]{|c|c|c|}
  \hline
\hspace{8mm} $D$\hspace{8mm} & \hspace{10mm} $E$\hspace{10mm}
& \hspace{8mm} $M$\hspace{8mm} \\
  \hline
    $ 3 $ & -0.5365 & 0.249 \\
  \hline
    $ 4 $ & -0.5456 & 0.228 \\
  \hline
    $ 5 $ & -0.5488 & 0.220 \\
  \hline
    $ 6 $ & -0.5513 & 0.206 \\
  \hline
    $ 7 $ & -0.5490 & 0.216 \\
  \hline
    $ 8 $ & -0.5506 & 0.212 \\
  \hline
  \end{tabular}
  \label{tab:energy}
\end{table}

\begin{table}[h]
  \centering
  \caption{Comparison of our results with those obtained by the other approaches for
  the ground state energy per site $E$ and the staggered magnetization $M$
  of the Heisenberg model with $h=0$.}
  \vspace{2mm}
  \begin{tabular}[b]{|c|c|c|}
  \hline
    \hspace{10mm} Method\hspace{10mm} & \hspace{10mm} $E$\hspace{10mm}
    & \hspace{8mm} M \hspace{8mm} \\
  \hline
    Spin wave\cite{Zheng1991}  & -0.5489 & 0.24\\
  \hline
    Series expansion\cite{Oitmaa1992} & -0.5443 & 0.27 \\
  \hline
    Monte Carlo\cite{Reger1989} & -0.5450 & 0.22 \\
  \hline
    Ours D=8 & -0.5506 & 0.21 $\pm$ 0.01 \\
  \hline
  \end{tabular}
  \label{tab:compare}
\end{table}


\begin{thebibliography}{99}

\bibitem{White1992} S. R. White, Phys. Rev. Lett. \textbf{69}, 2863 (1992).

\bibitem{Bursill96} R. J. Bursill, T. Xiang, and G. A. Gehring,
J. Phys.: Condens. Matt. \textbf{8} L583 (1996).

\bibitem{Wang97} X. Wang and T. Xiang, Phys. Rev. B \textbf{56}
5061 (1997).

\bibitem{Shollwock} U. Schollwock, Rev. Mod. Phys. \textbf{77},
259 (2005).

\bibitem{Xiang01} T. Xiang, J. Lou, and Z. B. Su,
Phys. Rev. B \textbf{64} 104414 (2001).


\bibitem{Xiang96} T. Xiang, Phys. Rev. B \textbf{53} R10445 (1996).


\bibitem{Ostlund1995} S. $\rm{\ddot{O}}$stlund and S. Rommer,
Phys. Rev. Lett. \textbf{75}, 3537 (1995).


\bibitem{Verstraete} F. Verstraete and J. Cirac, cond-mat/0407066.


\bibitem{Vidal2003} G. Vidal, Phys. Rev. Lett. \textbf{91}, 147902 (2003);
\textbf{93}, 040502 (2004); S. R. White and A. E. Feiguin, Phys.
Rev. Lett. \textbf{93}, 076401 (2004); A. J. Daley $et$ $al$., J.
Stat. Mech. (2004) P04005.

\bibitem{Vidal2007} G. Vidal, Phys. Rev. Lett. \textbf{98}, 070201
(2007).

\bibitem{Levin2007} Michael Levin and Cody P. Nave, Phys. Rev. Lett.
\textbf{99}, 120601 (2007).

\bibitem{Zheng1991} Zheng Weihong, J. Oitmaa, and C. J. Hamer, Phys. Rev. B \textbf{44}, 11869
(1991).

\bibitem{Oitmaa1992} J. Otimaa, C. J. Hamer, and Zheng Weihong,
Phys. Rev. B \textbf{45}, 9834 (1992).

\bibitem{Reger1989} J.D. Reger, J.A. Riera, and A.P. Young, J. Phys. C \textbf{1}, 1855
(1989).


\end{thebibliography}
\end{document}